\documentclass[twocolumn,twoside,superscriptaddress,prl,aps]{revtex4}
\usepackage{amsmath,mathrsfs,amsbsy,color,graphicx,bm,amsthm,amsfonts}
\usepackage{units}
\usepackage{bbm}
\usepackage{color}

\newcommand{\idol}{\ensuremath{\mathbbm 1}}

\newcommand{\tr}{{\rm Tr}}

\def\Re{\mathop{\rm Re}}
\def\Im{\mathop{\rm Im}}

\begin{document}

\title{Detecting and estimating continuous-variable entanglement by local orthogonal observables}
\author{Chengjie Zhang}
\affiliation{Centre for Quantum Technologies, National University of Singapore, 3 Science Drive 2, Singapore 117543, Singapore}
\author{Sixia Yu}
%\email{cqtys@nus.edu.sg}
\affiliation{Centre for Quantum Technologies, National University of Singapore, 3 Science Drive 2, Singapore 117543, Singapore}
\affiliation{
Hefei National Laboratory for Physical Sciences at Microscale and Department of Modern Physics,  University of Science and Technology of China, Hefei, Anhui 230026, China}
\author{Qing Chen}
\affiliation{Centre for Quantum Technologies, National University of Singapore, 3 Science Drive 2, Singapore 117543, Singapore}
\author{C.H. Oh}
%\email{phyohch@nus.edu.sg}
\affiliation{Centre for Quantum Technologies, National University of Singapore, 3 Science Drive 2, Singapore 117543, Singapore}
\affiliation{Physics Department, National University of Singapore, 3 Science Drive 2, Singapore 117543, Singapore}

\begin{abstract}
Entanglement detection and estimation are fundamental problems in quantum information science. Compared with discrete-variable states, for which lots of efficient entanglement detection criteria and lower bounds of entanglement measures have been proposed, the continuous-variable entanglement is much less understood. Here we shall present a family of entanglement witnesses based on continuous-variable local orthogonal observables (CVLOOs) to detect and estimate entanglement of Gaussian and non-Gaussian states, especially for bound entangled states. By choosing an optimal set of CVLOOs our entanglement witness is equivalent to the realignment criterion and can be used to detect bound entanglement of a class of $2+2$ mode Gaussian states. Via our entanglement witness, lower bounds of two typical entanglement measures for arbitrary two-mode continuous-variable states are provided.
\end{abstract}
\date{\today}


\maketitle
Entanglement is recognized as a valuable resource in quantum information processing. However, it is far from simple to determine whether or not a given state is entangled and how much entanglement it contains if the given state is indeed entangled, in both discrete-variable systems and continuous-variable systems. Therefore, entanglement detection and estimation are fundamental problems in quantum information theory \cite{review1}.

For entanglement detection, many efficient criteria have been proposed \cite{Peres,LUR1,CM,CCN,YU,nonlinear,optimal,
Simon,Duan,c4gs,c4Nha3,c4Nha4,c4Hillery,Shchukin,Nha,zhang,DiGuglielmo,Steinhoff}. In discrete variable systems, the famous positive partial transposition (PPT) criterion is necessary and sufficient in two-qubit and qubit-qutrit systems, but it is only necessary for separability in higher-dimensional systems \cite{Peres}. There exist entangled states with PPT known as bound entangled states for which many criteria \cite{LUR1,CM,CCN,YU,nonlinear,optimal} have also been proposed. For instance, the realignment criterion introduced in Ref. \cite{CCN} can be used to detect bound entanglement. Ref. \cite{YU} proposed a family of entanglement witnesses and corresponding positive maps that are not completely positive based on local orthogonal observables (LOOs), which can detect two kinds of bound entangled states. In continuous variable systems, although  a lot of entanglement criteria have been proposed for continuous variables, many of them are corollaries of the PPT criterion, or equivalent to the PPT criterion. For example, the entanglement conditions in Refs. \cite{Simon,Duan,c4gs,c4Nha3,c4Nha4,c4Hillery} are corollaries of the PPT criterion, and the infinite series of inequalities in Refs. \cite{Shchukin,Nha} are equivalent to the PPT criterion. Therefore, only a few criteria can be used to detect bound entangled states in continuous variable systems \cite{zhang,DiGuglielmo,Steinhoff}, and one needs more entanglement conditions of continuous variables to complement the PPT criterion.

For entanglement estimation, much interest has recently been focused on lower bounds of entanglement measures. Generally speaking, calculations of entanglement measures are formidable as the Hilbert space dimension increases. Up to now, only a few analytical results for certain entanglement measures have been derived, such as entanglement of formation (EOF) of two-qubit states \cite{2qubit1,2qubit2}, isotropic states \cite{eof1}, Werner states \cite{werner}, and two-mode symmetric Gaussian states \cite{gaussian}. In order to estimate entanglement, several lower bounds of entanglement measures have been proposed for discrete variables \cite{kai,mintert04,real,estimation,witness1,witness2}. However, unlike discrete variables there are few results about lower bounds of entanglement measures presented for continuous variables \cite{bound1,bound2}.

Our purpose in this work is two-fold: on the one hand, to construct an entanglement criterion which can detect continuous-variable bound entangled states; on the other hand, to propose lower bounds of entanglement measures for continuous-variable states. To this aim, we present a family of entanglement witnesses (EWs) based on LOOs for continuous-variable systems. The witnesses can detect entangled Gaussian and non-Gaussian states. Furthermore, we present the realignment criterion for Gaussian states which is equivalent to our entanglement witness with optimal choice of continuous-variable local orthogonal observables (CVLOOs). Using the realignment criterion, we detect bound entanglement in a class of $2+2$ mode Gaussian state. For entanglement estimation, lower bounds of entanglement measures for arbitrary two-mode continuous-variable states are proposed.

\textit{Continuous-variable local orthogonal observables.--}
The main tool we use is the CVLOOs which is a natural generalization of LOOs for discrete variables. In discrete variable systems, take a $d_A\times d_B$ system as an example, LOOs $\{G_k^\nu\}$  ($\nu$=$A$ or $B$) are a complete set of orthogonal bases of the observable space for subsystem $\nu$ \cite{YU}, which consists of $d_\nu^2$ observables satisfying $\tr(G_k^\nu G_l^\nu)=\delta_{kl}$ and $\varrho^\nu=\sum_{k=1}^{d_\nu^2}\tr(\varrho^\nu G_k^\nu)G_k^\nu$. The LOOs have been widely used in many problems for discrete variables, such as detecting bound entangled states \cite{YU}, necessary and sufficient condition for nonzero quantum discord \cite{condition1}, and quantifying quantum uncertainty based on skew information \cite{luo3}. However, to our knowledge there is no analogy of LOOs in continuous variable systems, and we present here the CVLOOs for the first time.

Let us focus on two-mode continuous-variable states, a complete set of CVLOOs  for each mode consists of infinite observables $G(\lambda)$ of this mode satisfying orthogonal relations
\begin{eqnarray}
\tr[G(\lambda)G(\lambda')]=\delta^{(2)}(\lambda-\lambda'),
\end{eqnarray}
and complete-set condition $\varrho=\int\langle G(\lambda)\rangle_{\varrho}G(\lambda)\ \mathrm{d}^2\lambda$, where $\lambda$ is a complex number index.
There are infinite complete sets of CVLOOs. For later use, we introduce one typical complete set of CVLOOs:
$\{\mathcal{G}(\lambda)\}=\{D(0)/\sqrt{\pi},[D(\alpha)+D^\dag(\alpha)]/\sqrt{2\pi},-i[D(\alpha)-D^\dag(\alpha)]/\sqrt{2\pi}\}$, where $D(\alpha)$ is
the Weyl displacement operator defined as $D(\alpha)=\exp(\alpha a^\dag-\alpha^* a)$, with $\alpha$ satisfying (i) $\Re\alpha>0$ or (ii) $\Re\alpha=0$ and $\Im\alpha>0$.

\textit{Detecting continuous variable entanglement by CVLOOs.--}
If we choose two arbitrary complete sets of CVLOOs $\{G(\lambda)\}$ for subsystem $A$ and $\{\widetilde{G}(\lambda)\}$ for subsystem $B$, respectively, we can construct the following EW candidate:
\begin{eqnarray}
    W&=&\idol-\int G(\lambda)\otimes \widetilde{G}(\lambda)\mathrm{d}^2\lambda.\label{EW}
\end{eqnarray}
That is because for any pure product state $\varrho=|a\rangle\langle a|\otimes |b\rangle\langle b|$ we have
$\tr (\varrho W)=1-\int \langle G(\lambda)\rangle_a \langle \widetilde{G}(\lambda)\rangle_b \mathrm{d}^2 \lambda\geq1-[\int \langle G(\lambda)\rangle^2_a \mathrm{d}^2 \lambda \int \langle \widetilde{G}(\lambda')\rangle^2_b \mathrm{d}^2 \lambda']^{1/2} =0$, where we have used the Cauchy inequality and the complete-set condition of CVLOOs. For any separable state $\tr (\varrho W)\geq 0$ holds because of the linearity.
Furthermore, we can associate a positive map to each EW candidate $W$ through the Jamio{\l}kowski isomorphism \cite{iso} as $\mathcal{O}(\varrho)=\tr_B(\idol\otimes \varrho^{T}W)=\idol\tr\varrho-\int\langle\widetilde{G}(\lambda)^{T}\rangle_{\varrho}G(\lambda)\mathrm{d}^2\lambda$. Therefore, we have an entanglement criterion based on the positive map: if a state $\varrho$ is separable then $\mathcal{O}\otimes\idol(\varrho)\geq0$,
where $\mathcal{O}\otimes\idol(\varrho)=\tr_A\varrho-\int\langle G(\lambda)\otimes \widetilde{G}(\lambda')\rangle_{\varrho}G_O(\lambda)\otimes \widetilde{G}(\lambda')\mathrm{d}^2\lambda\mathrm{d}^2\lambda'$, $\{G_O(\lambda)\}$ is another complete set of CVLOOs for subsystem $A$. If $G_O(\lambda)$ is the same as $G(\lambda)$, then we obtain the continuous-variable version of reduction criterion \cite{reduction}.

In the following, we construct a typical EWs belonging to Eq. (\ref{EW}), and calculate its expectation values for arbitrary two-mode state. The typical EW is as follows,
\begin{eqnarray}
    W_{\mu_1\mu_2}&=&\idol-\sqrt{|\mu_-\mu_+|}\int \mathcal{G}(\lambda)\otimes \mathcal{G}(-\mu_1\lambda-\mu_2\lambda^{*})\mathrm{d}^2\lambda,\ \ \ \ \  \label{EW1}
\end{eqnarray}
where we have used the Weyl displacement operator $D(\lambda)$ and  $\mu_{\pm}=\mu_1\pm\mu_2\neq0$ with $\mu_1,\mu_2$ being real parameters. For an arbitrary two-mode state $\varrho$, its characteristic function is defined as the expectation value of the two-mode Weyl displacement operator $\chi(\lambda_1,\lambda_2)=\tr[\varrho D_1(\lambda_1)D_2(\lambda_2)]$, and its Wigner function is defined as the Fourier transform of the characteristic function $W(\alpha_1,\alpha_2)=\pi^{-4}\int \exp[\sum_{i=1}^{2}(\lambda_i^*\alpha_i-\lambda_i\alpha_i^*)]\chi(\lambda_1,\lambda_2)\mathrm{d}^2\lambda_1\mathrm{d}^2\lambda_2$.
After some algebra, one can get the expectation value of $W_{\mu_1\mu_2}$,
\begin{equation}\label{EW1wigner}
\tr(\varrho W_{\mu_1\mu_2})=1-\pi\sqrt{|\mu_-\mu_+|}\int W(\mu_2\alpha^*-\mu_1\alpha,\alpha)\mathrm{d}^2\alpha.
\end{equation}
When $\tr(\varrho W_{\mu_1\mu_2})<0$, it is immediately indicated that $\varrho$ is entangled.

Let us define the position and momentum operators as $\hat{x}=(\hat{a}+\hat{a}^\dag)/2$ and $\hat{p}=-i(\hat{a}-\hat{a}^\dag)/2$, respectively.  The Wigner function of two-mode Gaussian states is: $W(\xi)=(2\pi)^{-2}(\mathrm{Det}\mathcal{V})^{-1/2}\exp(-\xi \mathcal{V}^{-1}\xi^{T}/2)$,
where the four-dimensional vector $\xi$ has the quadrature pairs of all two-modes as its components $\xi=(x_1,p_1,x_2,p_2)$ with $\hat{\xi}=(\hat{x}_1,\hat{p}_1,\hat{x}_2,\hat{p}_2)$, and $\mathcal{V}$ is the covariance matrix defined by $\mathcal{V}_{ij}=\tr[\varrho(\Delta\hat{\xi}_i\Delta\hat{\xi}_j+\Delta\hat{\xi}_j\Delta\hat{\xi}_i)/2]$ with $\Delta\hat{\xi}_i=\hat{\xi}_i-\langle\hat{\xi}_i\rangle$.
There is a standard form for the covariance matrix $\mathcal{V}$ of two-mode Gaussian state,
\begin{eqnarray}
\mathcal{V}=\left(\begin{array}{cc}
\mathcal{A}& \mathcal{C}\\
\mathcal{C}^{T}& \mathcal{B}
  \end{array}\right)\;,\label{CM}
\end{eqnarray}
where $\mathcal{A}=\mathrm{diag}(a,a)$, $\mathcal{B}=\mathrm{diag}(b,b)$ and $\mathcal{C}=\mathrm{diag}(c_1,c_2)$ with $a,b\geq1/4$ and $ab\geq c_1^2,c_2^2$. Under this standard form, one can arrive at $\tr(\varrho W_{\mu_1\mu_2})=1-\sqrt{|\mu_-\mu_+|}/[2\sqrt{(a+b\mu_{-}^2+2c_1\mu_{-})(a+b\mu_{+}^2+2c_2\mu_{+})}]$.
%\begin{equation}\label{gaussw1}
%    \tr(\varrho W_\mu)=1-\frac{\mu}{2\sqrt{(a\mu^2+b-2c_1\mu)(a\mu^2+b+2c_2\mu)}}.
%\end{equation}
It is obvious that the minimum of $\tr(\varrho W_{\mu_1\mu_2})$ is
\begin{eqnarray}
\tr(\varrho W_{\mu_1\mu_2})=1-\frac{1}{4\sqrt{(\sqrt{ab}-|c_1|)(\sqrt{ab}-|c_2|)}}\label{min}
\end{eqnarray}
for two-mode Gaussian states. Furthermore, this minimum EW condition is equivalent to the continuous variable PPT criterion shown in Ref. \cite{Simon} for all the symmetric two-mode Gaussian states. Since all the entangled two-mode Gaussian states can be transformed by local operations into symmetric entangled Gaussian states without destroying the entanglement \cite{Giedke}, one can use this EW with certain local operations to detect all the entangled two-mode Gaussian states. For symmetric two-mode non-Gaussian states, maybe the criteria shown in Ref. \cite{toth} can be generalized into continuous variable systems.

\begin{figure}
\includegraphics[scale=0.4]{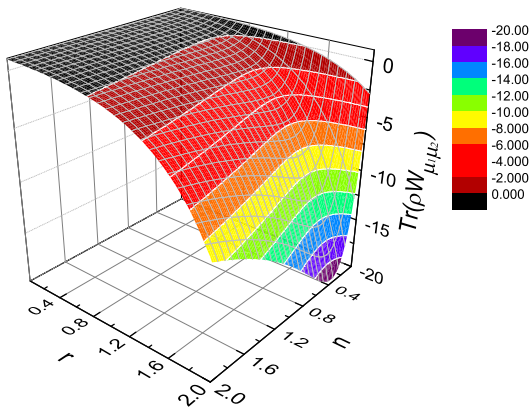}
\caption{(color online). Expectation value of the entanglement witness $W_{\mu_1\mu_2}$ shown in Eq. (\ref{EW1}) with  $\mu_1=0$ and $\mu_2=1$ for the single-photon-added two-mode symmetrical squeezed thermal state with parameters satisfying $n,r\in[0.02,2]$. All the states except in the black region can be detected by the entanglement witness. Moreover, one can also derive a simple lower bound of CREN for entangled states using Eq. (\ref{lowerbound}).}\label{1}
\end{figure}

The witness $W_{\mu_1\mu_2}$ can also be used for detecting entanglement of non-Gaussian states. For example, consider the single-photon-added two-mode symmetrical squeezed thermal state with its Wigner function given by
\begin{eqnarray}
W(\alpha_1,\alpha_2)&=&\frac{W_{STS}(\alpha_1,\alpha_2)}{(1 + 2 n)^2 (\cosh^2r + n \cosh2r)}\nonumber\\
&&\times\Big[(x_2 + 2 n x_2 + x_2\cosh2r-x_1 \sinh2r)^2 \nonumber\\
&&+ (p_2 + 2 n p_2 + p_2 \cosh2r + p_1 \sinh2r)^2\nonumber\\
&&-(1 + 2 n) (n + \cosh^2r)\Big],
\end{eqnarray}
%$W(\alpha_1,\alpha_2)=W_{STS}(\alpha_1,\alpha_2)[-(1 + 2 n) (n + \cosh^2r) + (x_2 + 2 n x_2 + x_2\cosh2r-x_1 \sinh2r)^2 + (p_2 + 2 n p_2 + p_2 \cosh2r + p_1 \sinh2r)^2]/[(1 + 2 n)^2 (\cosh^2r + n \cosh2r)]$
where $r$ and $n$ are the squeezing parameter and average photon number respectively, $\alpha_i=x_i+i p_i$ and $x_i$, $p_i$ are real parameters, and
%\begin{eqnarray}
%% \nonumber to remove numbering (before each equation)
%W_{STS}(\alpha_1,\alpha_2)&=&\exp\Bigg[\frac{-2(|\alpha_1|^2+|\alpha_2|^2) \cosh2r}{1+2n} \nonumber\\
%&&+\frac{4(x_1 x_2 - p_1 p_2) \sinh2r}{1 + 2 n}\Bigg]\frac{4}{(1 + 2 n)^2 \pi^2}\nonumber
%\end{eqnarray}
$W_{STS}(\alpha_1,\alpha_2)=4[(1 + 2 n) \pi]^{-2}\exp[-2(|\alpha_1|^2+|\alpha_2|^2) \cosh2r/(1 + 2 n) + 4(x_1 x_2 - p_1 p_2) \sinh2r/(1 + 2 n)]$
is the Wigner function of two-mode symmetrical squeezed thermal state. Using Eq. (\ref{EW1wigner}) with $\mu_1=0$ and $\mu_2=1$, one can derive $\tr(\varrho W_{\mu_1\mu_2})=1-e^{4 r} n (1+n)/[(1+2 n)^2 (\cosh^2r+n \cosh2r)]$.
Based on this expectation value, we have checked all the states with $n,r\in[0.02,2]$, and the results have been shown in Fig. \ref{1}. All the states except in the black region of Fig. \ref{1} have $\tr(\varrho W_{\mu_1\mu_2})<0$.

\textit{Realignment criterion and bound entanglement.--}
In discrete variable systems, the witness $W=\idol-\sum_k G_k^A\otimes G_k^B$ with optimal choice of LOOs is equivalent to the realignment criterion \cite{YU,nonlinear,optimal}, i.e., $\|\mathcal{R}(\varrho)\|\leq1$ for all separable states with $\mathcal{R}(A\otimes B):=A\otimes\idol(\sum_{ij}|ii\rangle\langle jj|)B^T\otimes\idol$ \cite{CCN}. The same situation exists for Eq. (\ref{EW}). Consider an $n+n$ mode state $\varrho$, which can be generally expressed by $\varrho=\pi^{-2n}\int W(\alpha_1,\cdots,\alpha_{2n})\exp[\sum_{i=1}^{2n}(\lambda_i^*\alpha_i-\lambda_i\alpha_i^*)]\otimes_{i=1}^{2n}D(\lambda_i)\prod_{i=1}^{2n}\mathrm{d}^2\alpha_i\mathrm{d}^2\lambda_i$. It is worth noticing that $\mathcal{R}(\varrho)=\pi^{-3n}\int W(\alpha_1,\cdots,\alpha_{2n})\exp[\sum_{i=1}^{2n}(\lambda_i^*\alpha_i-\lambda_i\alpha_i^*)][\otimes_{j=1}^n D(\lambda_j)D(\beta_j)D^T(\lambda_{n+j})][\otimes_{j=1}^nD(\beta_{j}^*)] \prod_{j=1}^{n}\mathrm{d}^2\beta_j \\
\prod_{i=1}^{2n}\mathrm{d}^2\alpha_i\mathrm{d}^2\lambda_i$. Therefore, the characteristic function of $\mathcal{R}(\varrho)\mathcal{R}^\dag(\varrho)$ is
$\chi(\mu_1,\cdots,\mu_{2n})=\pi^{2n}\int W(\alpha_1,\cdots,\alpha_{2n}) W(\alpha_1+(\mu_1+\mu_{n+1}^*)/2,\cdots,\alpha_n+(\mu_n+\mu_{2n}^*)/2,\alpha_{n+1},\cdots,\alpha_{2n})\prod_{j=1}^n \exp[(\mu_{n+j}-\mu_j^*)\alpha_j-(\mu_{n+j}^*-\mu_j)\alpha_j^*+(\mu_j\mu_{n+j}-\mu_j^*\mu_{n+j}^*)/2]\prod_{i=1}^{2n}\mathrm{d}^2\alpha_i$.
%\begin{widetext}
%\begin{eqnarray}
%% \nonumber to remove numbering (before each equation)
%\chi(\mu_1,\cdots,\mu_{2n})
%&=&\pi^{2n}\int W(\alpha_1,\cdots,\alpha_{2n}) W\bigg(\alpha_1+\frac{\mu_1+\mu_{n+1}^*}{2},\cdots,\alpha_n+\frac{\mu_n+\mu_{2n}^*}{2},\alpha_{n+1},\cdots,\alpha_{2n}\bigg)\nonumber\\
%&&\ \ \ \ \ \ \ \ \times\prod_{j=1}^n \exp[(\mu_{n+j}-\mu_j^*)\alpha_j-(\mu_{n+j}^*-\mu_j)\alpha_j^*]\exp\bigg(\frac{\mu_j\mu_{n+j}-\mu_j^*\mu_{n+j}^*}{2}\bigg)\prod_{i=1}^{2n}\mathrm{d}^2\alpha_i.
%\end{eqnarray}
%\end{widetext}
For Gaussian states, this characteristic function can also be written as $\chi(\mu_1,\cdots,\mu_{2n})=a_0\exp(-\Lambda \mathcal{V}_{\mathcal{R}\mathcal{R}^\dag}\Lambda^T/2)$ where $\Lambda=(b_1,a_1,\cdots,b_{2n},a_{2n})$ with $\mu_i=(a_i+ib_i)/2$, and $\mathcal{V}_{\mathcal{R}\mathcal{R}^\dag}$ is the covariance matrix of $\mathcal{R}(\varrho)\mathcal{R}^\dag(\varrho)$, hence one can obtain the covariance matrix $\mathcal{V}_{\mathcal{R}\mathcal{R}^\dag}$. According to the Williamson theorem \cite{williamson,vidal_adesso}, the covariance matrix $\mathcal{V}_{\mathcal{R}\mathcal{R}^\dag}$ can always be written in the diagonal form $\mathcal{V}_{\mathcal{R}\mathcal{R}^\dag}=S^T \mathbf{\nu} S$ where $S\in \mathrm{Sp}_{(4n,\mathbb{R})}$ is a symplectic transformation and $\mathbf{\nu}=\mathrm{diag}(\nu_1,\nu_1,\cdots,\nu_{2n},\nu_{2n})$ is the covariance matrix of a tensor product of thermal states. Thus, one can finally arrive at
\begin{eqnarray}
\|\mathcal{R}(\varrho)\|=\sqrt{a_0}\prod_{i=1}^{2n}\Big(\sqrt{2\nu_i+1/2}+\sqrt{2\nu_i-1/2}\Big).\label{R}
\end{eqnarray}
Note that Eq. (\ref{R}) can be calculated for all the $n+n$ mode Gaussian states. One only needs to get the characteristic function of $\mathcal{R}(\varrho)\mathcal{R}^\dag(\varrho)$ firstly, then get the covariance matrix $\mathcal{V}_{\mathcal{R}\mathcal{R}^\dag}$  and calculate its symplectic eigenvalues \cite{appendix}.

For two-mode Gaussian states with a standard form of covariance matrix in Eq. (\ref{CM}), the two symplectic eigenvalues are $\nu_i=\sqrt{ab}/(4\sqrt{ab-c_i^2})$ and $a_0=1/(16\prod_{i=1}^{2}\sqrt{ab-c_i^2})$. Therefore, one has $\|\mathcal{R}(\varrho)\|=1/(4\prod_{i=1}^{2}\sqrt{\sqrt{ab}-|c_i|})$ for two-mode Gaussian states which is exactly equivalent to Eq. (\ref{min}). It means the minimum of  $\tr(\varrho W_{\mu_1\mu_2})$ is realized under the optimal choice of CVLOOs.
It is worth noticing that Ref. \cite{xychen} has derived partial results of Eq. (\ref{min}), but no results about multi-mode Gaussian states until now.

For multi-mode states, consider the example of $2+2$ mode Gaussian state with its covariance matrix given by
\begin{eqnarray}
\mathcal{V}=\left(\begin{array}{cc}
a\idol_4& cR\\
cR^{T}& b\idol_4
  \end{array}\right)\ \mathrm{with}\
  R=\left(\begin{array}{cccc}
1& 0 & 0 & 0\\
0& 0 & 0 & -1\\
0& 0 & -1 & 0\\
0& -1 & 0 & 0
  \end{array}\right),
\end{eqnarray}
where $\idol_4$ is a $4\times4$ identity matrix, $a,b\geq1/4$ and $c$ is a real parameter. This covariance matrix corresponds to a valid state if and only if $|c|\leq\sqrt{ab-\sqrt{a^2+b^2-1/16}/4}$.  It can be checked that this $2+2$ mode Gaussian state is a PPT state, i.e., its partial transposition is still a valid state which means the PPT criterion is of no use. Moreover, if the state is detected as entangled state it must be bound entangled state. After some algebra, we can find its four symplectic eigenvalues which are the same $\nu_i=\sqrt{ab}/(4\sqrt{ab-c^2})$ and $a_0=1/[16(ab-c^2)]^2$. Therefore, based on Eq. (\ref{R}), the realignment criterion is that
\begin{eqnarray}
\|\mathcal{R}(\varrho)\|=\frac{1}{16(ab+c^2-2\sqrt{ab}|c|)}\leq1
\end{eqnarray}
holds for arbitrary separable states. \textit{Therefore, if $\|\mathcal{R}(\varrho)\|>1$, i.e., $\sqrt{ab}-1/4<|c|\leq\sqrt{ab-\sqrt{a^2+b^2-1/16}/4}$, the state must be bound entangled. For the special case $a=b$, the state reduces to Simon state \cite{simon2}, in which the realignment criterion $\|\mathcal{R}(\varrho)\|\leq1$ is not only a necessary condition but also a sufficient condition for separability}.

\textit{Estimating continuous variable entanglement.--}
Before embarking on our results, let us introduce two entanglement measures first. One famous entanglement measure is EOF. For pure state $|\varphi\rangle$, it is defined  by $E_F(|\varphi\rangle)=S(\varrho_A)$, where $S(\varrho)$ is the von Neumann entropy and $\varrho_A$ is the reduced density matrix of subsystem $A$. For mixed state $\varrho$, the EOF is defined by the convex roof, $E_F(\varrho)=\min_{\{p_i,|\varphi_i\rangle\}}\sum_i p_i E_F(|\varphi_i\rangle)$, where the minimum is taken over all possible ensemble realizations of $\varrho=\sum_i p_i |\varphi_i\rangle\langle\varphi_i|$. The other entanglement measure is convex-roof extended negativity (CREN). For pure state $|\varphi\rangle$, it is defined by the negativity $\mathcal{N}(|\varphi\rangle)=\||\varphi\rangle\langle\varphi|^{T_B}\|-1$ where $\|\cdot\|$ stands for the trace norm. For mixed states, CREN is also defined by the convex roof.

When $\mu_1=0$ and $\mu_2=1$, the entanglement witness $W_{\mu_1\mu_2}$ can not only be used for the detection of entanglement, but also for its quantification of CREN. It is worth noticing that for an arbitrary pure state $|\psi\rangle=U_A\otimes U_B \sum_k \sqrt{\mu_k}|kk\rangle$ with $\sqrt{\mu_k}$ being its Schmidt coefficients, we have $\mathcal{N}(|\psi\rangle)=(\sum_k\sqrt{\mu_k})^2-1$ and $\langle\psi|W_{\mu_1\mu_2}|\psi\rangle=1-|\sum_k\sqrt{\mu_k}\langle k|U_B^T U_A|k\rangle|^2$. Suppose that the minimal ensemble realization for $\mathcal{N}(\varrho)$ is $\{q_i,|\phi_i\rangle\}$. Therefore, one has a simple lower bound of CREN,
\begin{eqnarray}
\mathcal{N}(\varrho)&=&\min_{\{p_i,|\varphi_i\rangle\}}\sum_i p_i \mathcal{N}(|\varphi_i\rangle)=\sum_i q_i\mathcal{N}(|\phi_i\rangle)\nonumber\\
&\geq&-\sum_i q_i\langle\phi_i|W_{\mu_1\mu_2}|\phi_i\rangle=-\tr(\varrho W_{\mu_1\mu_2}),\label{lowerbound}
\end{eqnarray}
where the inequality holds since we have used the fact $(\sum_k\sqrt{\mu_k})^2\geq|\sum_k\sqrt{\mu_k}\langle k|U_B^T U_A|k\rangle|^2$. For example, one can get a lower bound of CREN for the single-photon-added symmetrical squeezed thermal state according to Eq. (\ref{lowerbound}):   $\mathcal{N}(\varrho)\geq e^{4 r} n (1+n)/[(1+2 n)^2 (\cosh^2r+n \cosh2r)]-1$.

The continuous-variable SWAP operator can be written as,
\begin{equation}\label{EW3}
    V=\int \mathcal{G}(\lambda)\otimes \mathcal{G}(\lambda)\mathrm{d}^2\lambda,
\end{equation}
which has the swapping property $V|\psi_1\rangle|\psi_2\rangle=|\psi_2\rangle|\psi_1\rangle$ with $|\psi_1\rangle$ and $|\psi_2\rangle$ being two arbitrary one-mode state. Thus, for an arbitrary two-mode separable state $\varrho_s$ one has $\tr(\varrho_s V)\geq0$ because of the swapping property. Similar to the EW $W_{\mu_1\mu_2}$, one can derive the expectation value of $V$ for an arbitrary two-mode state, $\tr(\varrho V)=\pi\int W(\alpha,\alpha)\mathrm{d}^2\alpha$,
where $\varrho$ is entangled provided $\tr(\varrho V)<0$. Interestingly, it can give a lower bound of EOF for the two-mode state $\varrho$ satisfying $\tr(\varrho V)<0$,
   \begin{eqnarray}
E_F(\varrho)\geq
H_2\bigg(\frac{1+\sqrt{1-(\tr\varrho V)^2}}{2}\bigg), \label{bound}
   \end{eqnarray}
where $H_2$ denotes the binary entropy function.
In order to get Eq. (\ref{bound}), we first prove that $C(\varrho)\geq-\tr(\varrho V)$, where $C(\varrho)$ denotes the concurrence of $\varrho$ defined by $C(|\varphi\rangle)=[2(1-\tr\varrho_A^2)]^{1/2}$ for pure states and convex roof for mixed states. Consider an arbitrary pure state $|\psi\rangle=U_A\otimes U_B \sum_k \sqrt{\mu_k}|kk\rangle$ with $\sqrt{\mu_k}$ being its Schmidt coefficients, we have $C(|\psi\rangle)=(2\sum_{k\neq k'}\mu_k\mu_{k'})^{1/2}$ and $\langle\psi|V|\psi\rangle=\sum_{kk'}\sqrt{\mu_k\mu_{k'}}\langle k'k'|U_A^\dag U_B\otimes U_B^\dag U_A|kk\rangle$.  Suppose that the minimal ensemble realization for $C(\varrho)$ is $\{q_i,|\phi_i\rangle\}$. Therefore, one has a simple lower bound of concurrence,
\begin{eqnarray}
C(\varrho)&=&\min_{\{p_i,|\varphi_i\rangle\}}\sum_i p_i C(|\varphi_i\rangle)=\sum_i q_iC(|\phi_i\rangle)\nonumber\\
&\geq&-\sum_i q_i\langle\phi_i|V|\phi_i\rangle=-\tr(\varrho V),
\end{eqnarray}
where the inequality holds since we have used the fact $(2\sum_{k\neq k'}\mu_k\mu_{k'})^{1/2}\geq-\sum_{kk'}\sqrt{\mu_k\mu_{k'}}\langle k'k'|U_A^\dag U_B\otimes U_B^\dag U_A|kk\rangle$
for arbitrary unitary matrices $U_A$ and $U_B$. It is worth noticing that one can acquire a lower bound of EOF from concurrence, i.e., $E_F(\varrho)\geq\mathrm{co}(R_L^{(n)}(C(\varrho)))$ where the function $\mathrm{co}(R_L^{(n)}(x))$ is a monotonically increasing function and $\mathrm{co}(R_L^{(n)}(x))=H_2(1/2+\sqrt{1-x^2}/2)$ when $0\leq x\leq1$ for arbitrary integer $n\geq2$ \cite{bound}. Therefore, Eq. (\ref{bound}) can be derived.

As the last example, consider the pure state $|\varphi\rangle=(|\alpha_1\rangle|\alpha_2\rangle-|\alpha_2\rangle|\alpha_1\rangle)/\sqrt{2}$ with vacuum-state noise, i.e. $\delta=p|\varphi\rangle\langle\varphi|+(1-p)|00\rangle\langle00|$,
where $|\alpha_1\rangle$ and $|\alpha_2\rangle$ are coherent states. With the swapping property, one can get that $\tr(\delta V)=p(\exp(-|\alpha_1-\alpha_2|^2)-1)+1-p$. When $p>1/[2-\exp(-|\alpha_1-\alpha_2|^2)]$, we have a lower bound of EOF for the state, $E_F(\delta)\geq H_2(1/2+\sqrt{1-(\tr\delta V)^2}/2)$.

\textit{Discussions and conclusions.--}
Some generalizations can be made for the above results. First of all, the entanglement witness Eq. (\ref{EW}) can be viewed as continuous-variable version of EW shown in Ref. \cite{YU}, and it can be improved to its nonlinear form, i.e., $F(\varrho)=1-\int [\langle G(\lambda)\otimes \widetilde{G}(\lambda)\rangle_{\varrho}+\langle G(\lambda)\otimes\idol - \idol\otimes\widetilde{G}(\lambda)\rangle^{2}_{\varrho}/2] \mathrm{d}^2\lambda$. For any pure product state $\varrho=|a\rangle\langle a|\otimes |b\rangle\langle b|$ we have $F(\varrho)=1-[\tr (|a\rangle\langle a|)^2 +\tr (|b\rangle\langle b|)^2]/2=0$. For any separable state $F(\varrho)\geq 0$ holds because of its concavity. This improved nonlinear form comes from the discrete-variable nonlinear EW given by Ref. \cite{nonlinear}, and can be regarded as its continuous-variable version as well.
Besides, it is worth noticing that the EW $W_\mu$ can be generalized as $W'=\idol-\int \mathcal{G}(\lambda)\otimes \mathcal{G}(f(\lambda))|\partial(\lambda_r,\lambda_i)/\partial(f_r(\lambda),f_i(\lambda))|^{-1/2}\mathrm{d}^2\lambda$, where $\lambda_r$ and $f_r(\lambda)$ ($\lambda_i$ and $f_i(\lambda)$) are real (imaginary) parts of $\lambda$ and $f(\lambda)$, respectively, and $\lambda\rightarrow f(\lambda)$ is bijective a map $\mathbb{C}\rightarrow\mathbb{C}$. Last but not least, from the entanglement witness $V$ one can provide lower bounds of other entanglement measures besides EOF and concurrence. For example, the tangle of $\varrho$ has a lower bound $\tau(\varrho)\geq[\tr(\varrho V)]^2$ when $\tr(\varrho V)<0$.

Besides the entanglement detection and estimation, the CVLOOs may have many other applications. For example, authors of Ref. \cite{condition1} have proposed a necessary and sufficient condition of nonzero quantum discord using LOOs for discrete variables. Using CVLOOs, one can also get a necessary and sufficient condition of nonzero quantum discord for continuous variables as well. Luo has introduced a measure quantifying quantum uncertainty based on skew information and LOOs \cite{luo3}, the measure can probably be extended to continuous variables using CVLOOs.  These potential applications will be of further research interest.

In conclusion, we present a family of entanglement witnesses based on LOOs for continuous-variable systems, which are used to detect entanglement of Gaussian and non-Gaussian states. We present the realignment criterion for Gaussian states which is equivalent to our entanglement witness with optimal choice of CVLOOs. Using the realignment criterion, we detect bound entanglement in a class of $2+2$ mode Gaussian state. Furthermore, lower bounds of entanglement measures for arbitrary two-mode continuous-variable states are also proposed.

We thank Otfried G\"uhne and Kazuo Fujikawa for discussions. This work is supported by the National Research Foundation and Ministry of Education, Singapore (Grant No. WBS: R-710-000-008-271), and the National Natural Science Foundation of China (Grant No. 11075227).

%\newpage
\setcounter{equation}{0}
\renewcommand{\theequation}{S\arabic{equation}}
\onecolumngrid

\section*{Supplemental Material}
Here we provide some details of the calculations. We have introduced one typical complete set of CVLOOs, which can be rewritten as
\begin{eqnarray}
\mathcal{G}(\lambda)=\left\{ \begin{array}{ll}
\frac{D(0)}{\sqrt{\pi}} & \textrm{if $\lambda=0$,}\\
\frac{D(\alpha)+D^\dag(\alpha)}{\sqrt{2\pi}} & \textrm{if $\Re\lambda>0$ or $\Re\lambda=0$ and $\Im\lambda>0$,}\\
\frac{D(\alpha)-D^\dag(\alpha)}{i\sqrt{2\pi}} & \textrm{if $\Re\lambda<0$ or $\Re\lambda=0$ and $\Im\lambda<0$,}
\end{array} \right.
\end{eqnarray}
where $D(\alpha)$ is the Weyl displacement operator defined as $D(\alpha)=\exp(\alpha a^\dag-\alpha^* a)$, with the complex number parameter $\alpha$ satisfying (i) $\Re\alpha>0$ or (ii) $\Re\alpha=0$ and $\Im\alpha>0$. The complete set of CVLOOs $\{\mathcal{G}(\lambda)\}$ comes from the non-Hermitian basis $\{D(\alpha)\}$. It can be checked that $\tr[\mathcal{G}(\lambda)\mathcal{G}(\lambda')]=\delta^{(2)}(\lambda-\lambda')$, since the Weyl displacement operator has the property $\tr[D(\alpha)D^\dag(\alpha')]=\pi\delta^{(2)}(\alpha-\alpha')$.

\section{A. Entanglement witness $W_{\mu_1\mu_2}$}
We have introduced a typical EW as follows,
\begin{eqnarray}
    W_{\mu_1\mu_2}&=&\idol-\sqrt{|\mu_-\mu_+|}\int \mathcal{G}(\lambda)\otimes \mathcal{G}(-\mu_1\lambda-\mu_2\lambda^{*})\mathrm{d}^2\lambda=\idol-\frac{\sqrt{|\mu_-\mu_+|}}{\pi}\int D(\lambda)\otimes D(\mu_1\lambda+\mu_2\lambda^{*})\mathrm{d}^2\lambda,\label{SEW1}
\end{eqnarray}
since $\tr[\mathcal{G}(-\mu_1\lambda-\mu_2\lambda^{*})\mathcal{G}(-\mu_1\lambda'-\mu_2\lambda'^{*})]=\delta^{(2)}[\mu_1(\lambda-\lambda')+\mu_2(\lambda^*-\lambda'^*)]=\delta^{(2)}(\lambda-\lambda')/|\mu_+\mu_-|$ and $\mu_-\mu_+\neq0$, Eq. (\ref{SEW1}) belongs to EW Eq. (2) in the main text, where $\mathcal{G}(\lambda)\otimes \mathcal{G}(-\mu_1\lambda-\mu_2\lambda^{*})$ denotes (i) $D(0)/\sqrt{\pi}\otimes D(0)/\sqrt{\pi}$ for $\lambda=0$; (ii) $[D(\alpha)+D^\dag(\alpha)]/\sqrt{2\pi}\otimes [D(-\mu_1 \alpha-\mu_2\alpha^*)+D^\dag(-\mu_1 \alpha-\mu_2\alpha^*)]/\sqrt{2\pi}$ for $\Re\lambda>0$ or $\Re\lambda=0$ and $\Im\lambda>0$; (iii) $[D(\alpha)-D^\dag(\alpha)]/(i\sqrt{2\pi})\otimes [D(-\mu_1 \alpha-\mu_2\alpha^*)-D^\dag(-\mu_1 \alpha-\mu_2\alpha^*)]/(i\sqrt{2\pi})$ for $\Re\lambda<0$ or $\Re\lambda=0$ and $\Im\lambda<0$, with $\alpha$ satisfying (i) $\Re\alpha>0$ or (ii) $\Re\alpha=0$ and $\Im\alpha>0$. Let us note that the characteristic function is defined as the expectation value of the two-mode Weyl displacement operator
\begin{eqnarray}
\chi(\lambda_1,\lambda_2)=\tr[\varrho_{AB} D_1(\lambda_1)D_2(\lambda_2)],
\end{eqnarray}
and its Wigner function is defined as the Fourier transform of the characteristic function
\begin{eqnarray}
W(\alpha_1,\alpha_2)=\frac{1}{\pi^{4}}\int \exp[\sum_{i=1}^{2}(\lambda_i^*\alpha_i-\lambda_i\alpha_i^*)]\chi(\lambda_1,\lambda_2)\mathrm{d}^2\lambda_1\mathrm{d}^2\lambda_2.
\end{eqnarray}
Conversely,
\begin{eqnarray}
\chi(\lambda_1,\lambda_2)&=&\int\exp[-\sum_{i=1}^{2}(\lambda_i^*\alpha_i-\lambda_i\alpha_i^*)]W(\alpha_1,\alpha_2)\mathrm{d}^2\alpha_1\mathrm{d}^2\alpha_2.
\end{eqnarray}
Therefore,
\begin{eqnarray}
\tr(\varrho W_{\mu_1\mu_2})&=&1-\frac{\sqrt{|\mu_-\mu_+|}}{\pi}\int \tr[D(\lambda)\otimes D(\mu_1\lambda+\mu_2\lambda^{*})]\mathrm{d}^2\lambda\nonumber\\
&=&1-\frac{\sqrt{|\mu_-\mu_+|}}{\pi}\int \chi(\lambda,\mu_1\lambda+\mu_2\lambda^{*})\mathrm{d}^2\lambda\nonumber\\
&=&1-\frac{\sqrt{|\mu_-\mu_+|}}{\pi}\int W(\alpha_1,\alpha_2) \exp(\lambda\alpha_1^*-\lambda^*\alpha_1)\exp[(\mu_1\lambda+\mu_2\lambda^{*})\alpha_2^*-(\mu_1\lambda^*+\mu_2\lambda)\alpha_2]\mathrm{d}^2\lambda\mathrm{d}^2\alpha_1\mathrm{d}^2\alpha_2\nonumber\\
&=&1-\frac{\sqrt{|\mu_-\mu_+|}}{\pi}\int W(\alpha_1,\alpha_2) \exp[\lambda^*(\mu_2\alpha_2^*-\mu_1\alpha_2-\alpha_1)-\lambda(\mu_2\alpha_2-\mu_1\alpha_2^*-\alpha_1^*)]\mathrm{d}^2\lambda\mathrm{d}^2\alpha_1\mathrm{d}^2\alpha_2\nonumber\\
&=&1-\frac{\sqrt{|\mu_-\mu_+|}}{\pi}\int W(\alpha_1,\alpha_2)\pi^2\delta^{(2)}(\mu_2\alpha_2^*-\mu_1\alpha_2-\alpha_1)\mathrm{d}^2\alpha_1\mathrm{d}^2\alpha_2\nonumber\\
&=&1-\pi\sqrt{|\mu_-\mu_+|}\int W(\mu_2\alpha_2^*-\mu_1\alpha_2,\alpha_2)\mathrm{d}^2\alpha_2,\label{SEWr1}
\end{eqnarray}
where we have used the identity $\int\exp(\lambda^*z-\lambda z^*)\mathrm{d}^2\lambda=\pi^2\delta^{(2)}(z)$.

Define the position and momentum operators as $\hat{x}=(\hat{a}+\hat{a}^\dag)/2$ and $\hat{p}=-i(\hat{a}-\hat{a}^\dag)/2$, respectively.  The Wigner function of the two-mode Gaussian state is:
\begin{equation}\label{}
    W(\alpha_1,\alpha_2)=\frac{1}{4\pi^2\sqrt{\mathrm{Det}\mathcal{V}}}\exp\Big(-\frac{1}{2}\xi \mathcal{V}^{-1}\xi^{\mathrm{T}}\Big),
\end{equation}
where $\mathcal{V}$ is the standard covariance matrix Eq. (5) in the main text, the four-dimensional vector $\xi=(x_1,p_1,x_2,p_2)$ and $\alpha_i=x_i+ip_i$. Substituting this Gaussian state Wigner function into Eq. (\ref{SEWr1}) and using the Gaussian function integral formula twice:
\begin{eqnarray}
\int_{-\infty}^{+\infty}\exp(-ax^2+bx+c)\mathrm{d}x=\sqrt{\frac{\pi}{a}}\exp\Bigg[\frac{b^2}{4a} + c\Bigg],  \ \ \ \ \ \mathrm{with}\ \  a>0,
\end{eqnarray}
one can get
\begin{eqnarray}
\tr(\varrho W_{\mu_1\mu_2})&=&1-\frac{\sqrt{|\mu_-\mu_+|}}{2\sqrt{(a+b\mu_{-}^2+2c_1\mu_{-})(a+b\mu_{+}^2+2c_2\mu_{+})}}\nonumber\\
&=&1-\frac{1}{2\sqrt{\Big(\frac{a}{|\mu_-|}+b|\mu_{-}|+2c_1\frac{\mu_{-}}{|\mu_-|}\Big)\Big(\frac{a}{|\mu_+|}+b|\mu_{+}|+2c_2\frac{\mu_{+}}{|\mu_+|}\Big)}}.
\end{eqnarray}
Since $a/|\mu_{\pm}|+b|\mu_{\pm}|\geq2\sqrt{ab}$, the equation holds when $|\mu_\pm|=\sqrt{a/b}$. In order to obtain the minimum of $\tr(\varrho W_{\mu_1\mu_2})$, the sign of $\mu_-$ and $\mu_+$ can be chosen so that $c_1\mu_-/|\mu_-|=-|c_1|$ and $c_2\mu_+/|\mu_+|=-|c_2|$, respectively.
Therefore, the minimum of $\tr(\varrho W_{\mu_1\mu_2})$ is
\begin{eqnarray}
\tr(\varrho W_{\mu_1\mu_2})=1-\frac{1}{4\sqrt{(\sqrt{ab}-|c_1|)(\sqrt{ab}-|c_2|)}}\label{smin}
\end{eqnarray}
when $\mu_-=-c_1\sqrt{a}/(|c_1|\sqrt{b})$ and $\mu_+=-c_2\sqrt{a}/(|c_2|\sqrt{b})$. For symmetric two-mode Gaussian states, i.e., $a=b$, this minimum EW condition is equivalent to the continuous variable PPT criterion shown by Simon \cite{ASimon}. Because Simon's PPT criterion for symmetric two-mode Gaussian states is that
\begin{eqnarray}
a^4+(\frac{1}{16}-|c_1c_2|)^2-a^2(c_1^2+c_2^2)\geq\frac{a^2}{8}
\end{eqnarray}
holds for arbitrary separable states, this condition is exactly equivalent to $\tr(\varrho W_{\mu_1\mu_2})\geq0$ in Eq. (\ref{smin}) with $a=b$.

\section{B. Realignment criterion for Gaussian states}
In discrete variable systems, the realignment criterion is $\|\mathcal{R}(\varrho)\|\leq1$ for all separable states with $\mathcal{R}(A\otimes B):=A\otimes\idol(\sum_{ij}|ii\rangle\langle jj|)B^T\otimes\idol$. In continuous variable system, consider an $n+n$ mode state $\varrho$, which can be generally expressed by
\begin{eqnarray}
\varrho=\frac{1}{\pi^{2n}}\int W(\alpha_1,\cdots,\alpha_{2n})\exp[\sum_{i=1}^{2n}(\lambda_i^*\alpha_i-\lambda_i\alpha_i^*)]\bigotimes_{i=1}^{2n}D(\lambda_i)\prod_{i=1}^{2n}\mathrm{d}^2\alpha_i\mathrm{d}^2\lambda_i.
\end{eqnarray}
Since in $n+n$ mode system
\begin{eqnarray}
 \sum_{ij}|ii\rangle\langle jj|=\frac{1}{\pi^n}\int [\bigotimes_{j=1}^n D(\beta_j)][\bigotimes_{j=1}^n D(\beta_{j}^*)]\prod_{j=1}^{n}\mathrm{d}^2\beta_j,
\end{eqnarray}
one can have
\begin{eqnarray}
\mathcal{R}(\varrho)=\frac{1}{\pi^{3n}}\int W(\alpha_1,\cdots,\alpha_{2n})\exp[\sum_{i=1}^{2n}(\lambda_i^*\alpha_i-\lambda_i\alpha_i^*)][\bigotimes_{j=1}^n D(\lambda_j)D(\beta_j)D^T(\lambda_{n+j})][\bigotimes_{j=1}^n D(\beta_{j}^*)] \prod_{j=1}^{n}\mathrm{d}^2\beta_j
\prod_{i=1}^{2n}\mathrm{d}^2\alpha_i\mathrm{d}^2\lambda_i.\nonumber\\
\end{eqnarray}
Therefore, the characteristic function of $\mathcal{R}(\varrho)\mathcal{R}^\dag(\varrho)$ is $\chi(\mu_1,\cdots,\mu_{2n})=\tr[\mathcal{R}(\varrho)\mathcal{R}^\dag(\varrho)\otimes_{i=1}^{2n} D(\mu_i)]$, i.e.,
\begin{eqnarray}
\chi(\mu_1,\cdots,\mu_{2n})
&=&\pi^{2n}\int W(\alpha_1,\cdots,\alpha_{2n}) W\bigg(\alpha_1+\frac{\mu_1+\mu_{n+1}^*}{2},\cdots,\alpha_n+\frac{\mu_n+\mu_{2n}^*}{2},\alpha_{n+1},\cdots,\alpha_{2n}\bigg)\nonumber\\
&&\ \ \ \ \ \ \ \ \times\prod_{j=1}^n \exp[(\mu_{n+j}-\mu_j^*)\alpha_j-(\mu_{n+j}^*-\mu_j)\alpha_j^*]\exp\bigg(\frac{\mu_j\mu_{n+j}-\mu_j^*\mu_{n+j}^*}{2}\bigg)\prod_{i=1}^{2n}\mathrm{d}^2\alpha_i.
\end{eqnarray}
For Gaussian states, this characteristic function is still a Gaussian function, and it can be written as
\begin{eqnarray}
\chi(\mu_1,\cdots,\mu_{2n})=a_0\exp(-\frac{1}{2}\Lambda \mathcal{V}_{\mathcal{R}\mathcal{R}^\dag}\Lambda^T),
\end{eqnarray}
where $\Lambda=(b_1,a_1,\cdots,b_{2n},a_{2n})$ with $\mu_i=(a_i+ib_i)/2$, and $\mathcal{V}_{\mathcal{R}\mathcal{R}^\dag}$ is the covariance matrix of $\mathcal{R}(\varrho)\mathcal{R}^\dag(\varrho)$, hence one can get the covariance matrix $\mathcal{V}_{\mathcal{R}\mathcal{R}^\dag}$. According to the Williamson theorem \cite{Awilliamson}, the covariance matrix $\mathcal{V}_{\mathcal{R}\mathcal{R}^\dag}$ can always be written in the diagonal form $\mathcal{V}_{\mathcal{R}\mathcal{R}^\dag}=S^T \mathbf{\nu} S$ where $S\in \mathrm{Sp}_{(4n,\mathbb{R})}$ is a symplectic transformation and $\mathbf{\nu}=\mathrm{diag}(\nu_1,\nu_1,\cdots,\nu_{2n},\nu_{2n})$ is the covariance matrix of a tensor product of thermal states given by
\begin{eqnarray}
\varrho_{\mathcal{R}\mathcal{R}^\dag}=\bigotimes_{i=1}^{2n}\frac{1}{\bar{n}_i +1}\sum_{k=0}^{\infty}\Big(\frac{\bar{n}_i}{\bar{n}_i +1}\Big)^k |k\rangle\langle k|,
\end{eqnarray}
where $\bar{n}_i=2\nu_i-1/2$ is the average photon number of each mode. Denote the matrix $J$ as
\begin{eqnarray}
J=\bigoplus_{i=1}^{2n}\left(\begin{array}{cc}
0& 1\\
-1& 0
  \end{array}\right)\;,
\end{eqnarray}
the fast way to compute $\nu_i$ is via the eigenvalues of the matrix $J^{-1}\mathcal{V}_{\mathcal{R}\mathcal{R}^\dag}$, which are $\pm i\nu_1,\cdots, \pm i\nu_{2n}$. Therefore,
\begin{eqnarray}
\|\mathcal{R}(\varrho)\|=\sqrt{a_0}\prod_{i=1}^{2n}\sum_{k=0}^{\infty} \frac{\bar{n}_i^{\frac{k}{2}}}{(\bar{n}_i +1)^{\frac{k+1}{2}}}.
\end{eqnarray}
Using the identity
\begin{eqnarray}
\sum_{k=0}^{\infty} \frac{\bar{n}_i^{\frac{k}{2}}}{(\bar{n}_i +1)^{\frac{k+1}{2}}}=\frac{1}{\sqrt{\bar{n}_i+1}-\sqrt{\bar{n}_i}},
\end{eqnarray}
one can finally arrive at
\begin{eqnarray}
\|\mathcal{R}(\varrho)\|=\sqrt{a_0}\prod_{i=1}^{2n}\Big(\sqrt{2\nu_i+1/2}+\sqrt{2\nu_i-1/2}\Big).\label{sR}
\end{eqnarray}
Note that Eq. (\ref{sR}) can be calculated for all the $n+n$ mode Gaussian states. One only needs to get the characteristic function of $\mathcal{R}(\varrho)\mathcal{R}^\dag(\varrho)$ firstly, then to get the covariance matrix $\mathcal{V}_{\mathcal{R}\mathcal{R}^\dag}$  and calculate its symplectic eigenvalues.

For examples, the covariance matrix $\mathcal{V}_{\mathcal{R}\mathcal{R}^\dag}$ of two-mode Gaussian states with a standard form of covariance matrix is
\begin{eqnarray}
\mathcal{V}_{\mathcal{R}\mathcal{R}^\dag}=\left(
\begin{array}{cccc}
 \frac{b+16 a (a b- c_2^2)}{32 ( a b- c_2^2)} & 0 & -\frac{b-16 a (a b- c_2^2)}{32 ( a b- c_2^2)} & 0 \\
 0 & \frac{b+16 a (a b- c_1^2)}{32 ( a b- c_1^2)} & 0 & \frac{b-16 a (a b- c_1^2)}{32 ( a b- c_1^2)} \\
 -\frac{b-16 a (a b- c_2^2)}{32 ( a b- c_2^2)} & 0 & \frac{b+16 a (a b- c_2^2)}{32 ( a b- c_2^2)} & 0 \\
 0 & \frac{b-16 a (a b- c_1^2)}{32 ( a b- c_1^2)} & 0 & \frac{b+16 a (a b- c_1^2)}{32 ( a b- c_1^2)}
\end{array}
\right).
\end{eqnarray}
The two symplectic eigenvalues are $\nu_i=\sqrt{ab}/(4\sqrt{ab-c_i^2})$ and $a_0=1/(16\prod_{i=1}^{2}\sqrt{ab-c_i^2})$. Therefore, one has
\begin{eqnarray}
\|\mathcal{R}(\varrho)\|=\frac{1}{4\prod_{i=1}^{2}\sqrt{\sqrt{ab}-|c_i|}}
\end{eqnarray}
for two-mode Gaussian states which is exactly equivalent to Eq. (\ref{smin}).
For higher-mode states, consider the example of $2+2$ mode Gaussian state with its covariance matrix given by
\begin{eqnarray}
\mathcal{V}=\left(\begin{array}{cc}
a\idol_4& cR\\
cR^{T}& b\idol_4
  \end{array}\right)\ \mathrm{with}\
  R=\left(\begin{array}{cccc}
1& 0 & 0 & 0\\
0& 0 & 0 & -1\\
0& 0 & -1 & 0\\
0& -1 & 0 & 0
  \end{array}\right),
\end{eqnarray}
where $\idol_4$ is a $4\times4$ identity matrix, $a,b\geq1/4$ and $c$ is a real parameter. This covariance matrix corresponds to a valid state if and only if $\mathcal{V}+i J/4\geq0$ i.e. $|c|\leq\sqrt{ab-\sqrt{a^2+b^2-1/16}/4}$.  It can be checked that this $2+2$ mode Gaussian state is a PPT state, i.e., its partial transposition is still a valid state which means the PPT criterion is of no use. Moreover, if the state is detected as entangled state it must be bound entangled state. After some algebra, we can find its covariance matrix $\mathcal{V}_{\mathcal{R}\mathcal{R}^\dag}$ as
\begin{eqnarray}
\left(
\begin{array}{cccccccc}
 \frac{b+16 a(ab-c^2)}{32 (a b-c^2)} & 0 & 0 & 0 & -\frac{b-16 a(ab-c^2)}{32( a b-c^2)} & 0 & 0 & 0 \\
 0 & \frac{b+16 a(ab-c^2)}{32( a b-c^2)} & 0 & 0 & 0 & \frac{b-16 a(ab-c^2)}{32( a b-c^2)} & 0 & 0 \\
 0 & 0 & \frac{b+16 a(ab-c^2)}{32( a b-c^2)} & 0 & 0 & 0 & -\frac{b-16 a(ab-c^2)}{32( a b-c^2)} & 0 \\
 0 & 0 & 0 & \frac{b+16 a(ab-c^2)}{32( a b-c^2)} & 0 & 0 & 0 & \frac{b-16 a(ab-c^2)}{32( a b-c^2)} \\
 -\frac{b-16 a(ab-c^2)}{32( a b-c^2)} & 0 & 0 & 0 & \frac{b+16 a(ab-c^2)}{32( a b-c^2)} & 0 & 0 & 0 \\
 0 & \frac{b-16 a(ab-c^2)}{32( a b-c^2)} & 0 & 0 & 0 & \frac{b+16 a(ab-c^2)}{32( a b-c^2)} & 0 & 0 \\
 0 & 0 & -\frac{b-16 a(ab-c^2)}{32( a b-c^2)} & 0 & 0 & 0 & \frac{b+16 a(ab-c^2)}{32( a b-c^2)} & 0 \\
 0 & 0 & 0 & \frac{b-16 a(ab-c^2)}{32( a b-c^2)} & 0 & 0 & 0 & \frac{b+16 a(ab-c^2)}{32( a b-c^2)}
\end{array}
\right).\nonumber
\end{eqnarray}
Its four symplectic eigenvalues are the same $\nu_i=\sqrt{ab}/(4\sqrt{ab-c^2})$ and $a_0=1/[16(ab-c^2)]^2$. Therefore, based on Eq. (\ref{sR}), the realignment criterion is that
\begin{eqnarray}
\|\mathcal{R}(\varrho)\|=\frac{1}{16(ab+c^2-2\sqrt{ab}|c|)}\leq1
\end{eqnarray}
holds for arbitrary separable states. If $\|\mathcal{R}(\varrho)\|>1$ i.e. $\sqrt{ab}-1/4<|c|\leq\sqrt{ab-\sqrt{a^2+b^2-1/16}/4}$, the state must be bound entangled. For the special case $a=b$, the state reduces to Simon state \cite{Asimon2}, in which the realignment criterion $\|\mathcal{R}(\varrho)\|\leq1$ is not only a necessary condition but also a sufficient condition for separability.

\section{C. Estimating continuous variable entanglement}
When $\mu_1=0$ and $\mu_2=1$, the entanglement witness $W_{\mu_1\mu_2}$ can be rewritten as
\begin{eqnarray}
W_{\mu_1\mu_2}=\idol-\int \mathcal{G}(\lambda)\otimes\mathcal{G}(-\lambda^*)\mathrm{d}^2\lambda=\idol-\frac{1}{\pi}\int D(\lambda)\otimes D(\lambda^*)\mathrm{d}^2\lambda=\idol-\sum_{ij}|ii\rangle\langle jj|.
\end{eqnarray}
It is worth noticing that for an arbitrary pure state $|\psi\rangle=U_A\otimes U_B \sum_k \sqrt{\mu_k}|kk\rangle$ with $\sqrt{\mu_k}$ being its Schmidt coefficients, we have $\mathcal{N}(|\psi\rangle)=(\sum_k\sqrt{\mu_k})^2-1$ and $\langle\psi|W_{\mu_1\mu_2}|\psi\rangle=1-|\sum_k\sqrt{\mu_k}\langle k|U_B^T U_A|k\rangle|^2$. Suppose that the minimal ensemble realization for $\mathcal{N}(\varrho)$ is $\{q_i,|\phi_i\rangle\}$. Therefore, one has a simple lower bound of CREN,
\begin{eqnarray}
\mathcal{N}(\varrho)=\min_{\{p_i,|\varphi_i\rangle\}}\sum_i p_i \mathcal{N}(|\varphi_i\rangle)=\sum_i q_i\mathcal{N}(|\phi_i\rangle)
\geq-\sum_i q_i\langle\phi_i|W_{\mu_1\mu_2}|\phi_i\rangle=-\tr(\varrho W_{\mu_1\mu_2}),\label{slowerbound}
\end{eqnarray}
where we have used the fact $(\sum_k\sqrt{\mu_k})^2\geq|\sum_k\sqrt{\mu_k}\langle k|U_B^T U_A|k\rangle|^2$ since $|\langle k|U_B^T U_A|k\rangle|\leq1$.

The continuous-variable SWAP operator can be written as,
\begin{equation}\label{sEW3}
    V=\int \mathcal{G}(\lambda)\otimes \mathcal{G}(\lambda)\mathrm{d}^2\lambda=\frac{1}{\pi}\int D(\lambda)\otimes D^\dag(\lambda)\mathrm{d}^2\lambda=\sum_{ij}|ij\rangle\langle ji|.
\end{equation}
Similar to the entanglement witness $W_{\mu_1\mu_2}$, one can get the expectation value of $V$ for an arbitrary two-mode state,
\begin{equation}\label{}
    \tr(\varrho V)=\pi\int W(\alpha,\alpha)\mathrm{d}^2\alpha,
\end{equation}
where $\varrho$ is entangled provided $\tr(\varrho V)<0$.  $C(\varrho)$ denotes the concurrence of $\varrho$ defined by $C(|\varphi\rangle)=[2(1-\tr\varrho_A^2)]^{1/2}$ for pure states and convex roof for mixed states. Consider an arbitrary pure state $|\psi\rangle=U_A\otimes U_B \sum_k \sqrt{\mu_k}|kk\rangle$ with $\sqrt{\mu_k}$ being its Schmidt coefficients, we have $C(|\psi\rangle)=(2\sum_{k\neq k'}\mu_k\mu_{k'})^{1/2}$ and $\langle\psi|V|\psi\rangle=\sum_{kk'}\sqrt{\mu_k\mu_{k'}}\langle k'k'|U_A^\dag U_B\otimes U_B^\dag U_A|kk\rangle$.  Suppose that the minimal ensemble realization for $C(\varrho)$ is $\{q_i,|\phi_i\rangle\}$. Therefore, one has a simple lower bound of concurrence,
\begin{eqnarray}
C(\varrho)=\min_{\{p_i,|\varphi_i\rangle\}}\sum_i p_i C(|\varphi_i\rangle)=\sum_i q_iC(|\phi_i\rangle)
\geq-\sum_i q_i\langle\phi_i|V|\phi_i\rangle=-\tr(\varrho V),
\end{eqnarray}
where the inequality holds since we have used the fact that
\begin{eqnarray}
\sqrt{2\sum_{k\neq k'}\mu_k\mu_{k'}}\geq-\sum_{kk'}\sqrt{\mu_k\mu_{k'}}\langle k'k'|U_A^\dag U_B\otimes U_B^\dag U_A|kk\rangle\label{sfact}
\end{eqnarray}
holds for arbitrary unitary matrices $U_A$ and $U_B$. Let us prove Eq. (\ref{sfact}). Define
\begin{eqnarray}
M_{kl}:=\frac{u_{kl}u_{lk}^*+u_{lk}u_{kl}^*}{2}
\end{eqnarray}
where $u_{kl}=\langle k|U_A^\dag U_B|l\rangle$, and define three sets $P_+=\{(k,l)|M_{kl}\geq0\}$, $P_-=\{(k,l)|M_{kl}<0\}$ and $P_-(k)=\{l|(k,l)\in P_-\}$.
Therefore,
\begin{eqnarray}
-\sum_{kk'}\sqrt{\mu_k\mu_{k'}}\langle k'k'|U_A^\dag U_B\otimes U_B^\dag U_A|kk\rangle=-\sum_{kk'}\sqrt{\mu_k\mu_{k'}}M_{kk'}\leq-\sum_{(k,k')\in P_-}\sqrt{\mu_k\mu_{k'}}M_{kk'}:=\Delta_-.
\end{eqnarray}
It is worth noticing that
\begin{eqnarray}
\Delta_-^2=\sum_{(k,k')\in P_- \atop (l,l')\in P_-}\sqrt{\mu_k\mu_{k'}\mu_l\mu_{l'}}M_{kk'}M_{ll'}=I_1+I_2+I_3+I_4,
\end{eqnarray}
where $I_1,I_2,I_3,I_4$ are corresponding to the cases: (i) $k=l$ and $k'=l'$ (ii) $k=l$ and $k'\neq l'$ (iii) $k\neq l$ and $k'=l'$ (iv) $k\neq l$ and $k'\neq l'$, respectively. Thus,
\begin{eqnarray}
I_1&=&\sum_{(k,k')\in P_-}\mu_k\mu_{k'}M_{kk'}^2, \\
I_2&=&\sum_{(k,k')\in P_-, \atop   (k,l')\in P_-, k'\neq l'}\mu_k\sqrt{\mu_{k'}\mu_{l'}}M_{kk'}M_{kl'}\leq \sum_{(k,k')\in P_-}\mu_k\mu_{k'}M_{kk'}\sum_{l'\in P_-(k) \atop l'\neq k'}M_{kl'}, \\
I_3&=&I_2, \\
I_1+I_2+I_3&\leq& \sum_{(k,k')\in P_-}\mu_k\mu_{k'}\Big( \sum_{l'\in P_-(k)}M_{kl'}  \Big)^2\leq\sum_{(k,k')\in P_-}\mu_k\mu_{k'}, \\
I_4&=&\sum_{(k,k')\in P_-,k\neq l \atop (l,l')\in P_-, k'\neq l'}\sqrt{\mu_k\mu_{k'}\mu_l\mu_{l'}}M_{kk'}M_{ll'}\leq \sum_{k\neq l}\mu_k\mu_l \sum_{k'\in P_-(k), l'\in P_-(l) \atop k'\neq l'} M_{kk'}M_{ll'}\nonumber\\
&\leq&  \sum_{k\neq l}\mu_k\mu_l \Big(\sum_{k'\in P_-(k)} M_{kk'}   \Big)^2\leq \sum_{k\neq l}\mu_k\mu_l,
\end{eqnarray}
where we have used $\sqrt{\mu_k\mu_l}\leq (\mu_k+\mu_l)/2$ and $\sum_{l}|M_{kl}|\leq\sum_l |u_{kl}|\cdot|u_{lk}|\leq1$. Therefore, one has
\begin{eqnarray}
\Delta_-^2\leq 2\sum_{k\neq l}\mu_k\mu_l,
\end{eqnarray}
from which we can derive Eq. (\ref{sfact}).

\end{document}